\documentclass[]{elsarticle}

\usepackage{amsfonts,graphicx}
\usepackage{bm}
\usepackage{hyperref,color}

\newcommand{\z}[1]{\boldsymbol{\mathrm{#1}}}

\begin{document}

\title{Stochastic Impedance}

\author{Bart Cleuren}
\ead{bart.cleuren@uhasselt.be}
\author{Karel Proesmans}
\ead{Karel.proesmans@uhasselt.be}
\address{Theoretical Physics, Hasselt University, B-3590 Diepenbeek, Belgium.}

\begin{abstract}
The concept of impedance, which characterises the current response to a periodical driving, is introduced in the context of stochastic transport. In particular, we calculate the impedance for an exactly solvable model, namely the stochastic transport of particles through a single-level quantum dot.
\end{abstract}


\maketitle

%\tableofcontents
\section{Introduction}
Electrical impedance \cite{young,horowitz} is a well known concept from undergraduate courses on electromagnetism. Students learn about it when analysing simple electrical circuits composed out of resistors, capacitors and inductors. When driven by an alternating voltage, the current response of these circuits is far richer compared to any direct current measurement. Apart from the amplitude, now also the phase shift between the voltage and current signal comes into play. By combining amplitude and phase into a single complex quantity, the impedance, students learn to appreciate the effectiveness of a complex notation. In a typical experimental setup, a swipe of the driving frequency allows access to detailed information about the composition and topology of the underlying electrical network.

Impedance can also be found in many other areas such as acoustics \cite{kinsler1999fundamentals} and mechanical response \cite{sabanovic2011motion}. All it requires is a linear and time-invariant system, properties which are easily found in various branches of physics. In this work, we introduce the concept of impedance in the context of stochastic transport of particles. We do so by considering one of the simplest stochastic systems available, namely a single-level quantum dot connected to two electron reservoirs. Periodic driving is achieved by modulating the difference in chemical potential between the reservoirs. Such a driving brings the system out of equilibrium and induces a flow of particles between the reservoirs via the intermediate quantum dot. As we demonstrate, even for this simple setup the response behaviour is rich and intricate. And quite remarkable, for certain parameter values the impedance of this system can be mapped exactly to the one of an equivalent electrical circuit.

The inspiration for this work comes from a series of remarkable experiments performed at the X-LAB facilities of Hasselt University \cite{cornelissen2018cell}. Subject of these investigations are cable bacteria, a recently discovered organism found in fresh and seawater sediments \cite{nielsen2010electric,pfeffer2012filamentous,meysman2015,meysman2017}. These multicellular organisms form centimeter long and unbranched filaments. Due to their specific living environment, they have developed a unique metabolism which requires them to transport electrons from one end of the filament to the other end. Recent experiments in which a voltage difference is applied across the filament, show that these cable bacteria are capable of conducting electrons over centimeter long distances, an organic tour de force. The mechanism by which they achieve this, however, remains elusive, and is the subject of ongoing research. One possibility is an incoherent hopping of electrons between discrete sites along the conductive pathway \cite{creasey2018}. In this context, the use of impedance spectroscopy is a well tried technique to further characterise the electrical conduction. In the present work, we investigate if and how such a stochastic hopping motion would be reflected in a measurement of the impedance. 

This paper is organised as follow. In section \ref{e_imp} we set the scene and briefly discuss the electrical impedance. This section introduces the basic notions, and the impedance of a simple electrical circuit is calculated. With hindsight, this impedance will be of relevance in section \ref{s_imp}. In that section, we introduce the concept of stochastic impedance by studying the stochastic transport through a quantum dot. Furthermore, we discuss the relation between stochastic and electrical impedance. In section \ref{outlook} we conclude and discuss several possibilities for future research.

%%%%%%%%%%%%%%%%%%%%%%%%%%%%%%%%%%%%%%%%%%% 
\section{Electrical Impedance}\label{e_imp}
\begin{figure}\begin{centering}
\includegraphics[width=11cm]{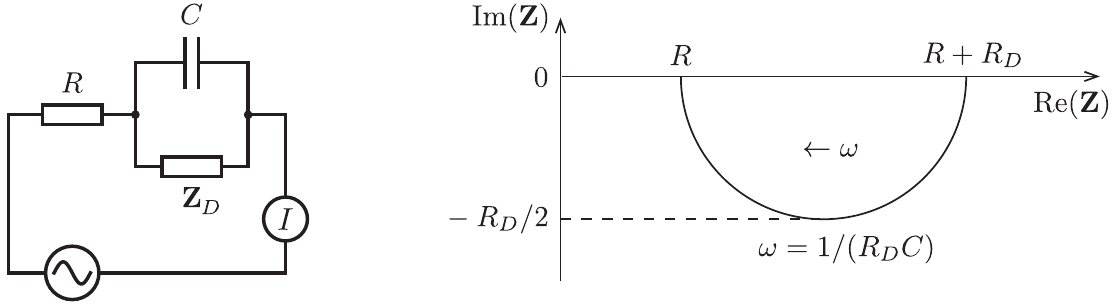}
\caption{(left panel) A standard electrical circuit pertaining to an impedance spectroscopy experiment, with the purpose to characterise the unknown impedance $\z{Z}_D$. $R$ and $C$ represent the resistance and capacity originating from the measurement setup. (right panel) Plot of the imaginary and real part of the impedance $\z{Z}$ of the circuit shown in the left panel, with $\z{Z}_D$ replaced by a (pure) resistor $R_D$.\label{fig1}}
\end{centering}\end{figure}

Consider an electrical circuit composed of passive and linear components, and driven by a sinusoidal voltage $V(t)=V_0 \cos(\omega t +\varphi)$ with frequency $\omega$.  The linear dependence between voltage and current implies that the current signal has the exact same frequency, albeit with a different phase. Combining amplitude $V_0$ and phase $\varphi$ of the voltage signal into a single complex quantity $\z{V}=V_0 e^{i\varphi}$, the applied voltage can be written as the following real part
\begin{equation}
    V(t)=\mathrm{Re}\left(\z{V} e^{i\omega t}\right),\label{Vdef}
\end{equation}
and a similar expression holding for the induced current 
\begin{equation}
    I(t)=\mathrm{Re}\left(\z{I} e^{i\omega t}\right).
\end{equation}
The impedance is then defined as the ratio of these complex amplitudes
\begin{equation}
    \z{Z}=\z{V}/\z{I}.
\end{equation}
For the basic electrical components this leads to:
\begin{equation}
\begin{array}{ll}
\z{Z}_R=R & \textrm{resistor $R$},\\
\z{Z}_C=1/(i\omega C) & \textrm{capacitor $C$},\\
\z{Z}_L=i\omega L & \textrm{inductor $L$}.
\end{array}
\end{equation}
A major advantage of the complex impedance is that the familiar rules apply for serial and parallel combination. As an example, consider the electric circuit given in Fig.~\ref{fig1}. This elementary circuit (and variants) is commonly found in impedance experiments \cite{impedance}. The resistor $R$ and capacitance $C$ can be seen as originating from the contact resistance and capacitance of the measurement setup, whereas $\z{Z}_D$ is the (unknown) impedance of the device of interest. The total impedance of this circuit follows immediately:
\begin{equation}
\z{Z}=R+\left(i\omega C+\frac{1}{\z{Z}_D}\right)^{-1}.
\end{equation}
The case of a purely resistive device, ie. for $\z{Z}_D\equiv R_D$, leads to
\begin{equation}
\z{Z}=R+\frac{R_D}{1+C^2 R^2_D \omega^2}-i\frac{\omega CR^2_D}{1+C^2 R^2_D \omega^2}. \label{zRD}
\end{equation}
A visualization of the impedance can be done by a parametric plot of the real and complex parts. In case of $\z{Z}_D\equiv R_D$, this results in the well-known semi-circle shown in the right panel of Fig.~\ref{fig1}. A quick calculation indeed confirms that 
\begin{equation}
    \left(\textrm{Re}(\z{Z})-R-\frac{R_D}{2}\right)^2+\textrm{Im}(\z{Z})^2=\frac{R_D^2}{4},\label{RIZ}
\end{equation}
hence the real and imaginary parts lie on a circle of radius $R_D/2$ with its centre on the real (horizontal) axis.

%%%%%%%%%%%%%%%%%%%%%%%%%%%%%%%%%%%%%%%%%%% 
\section{Stochastic Impedance: transport through a quantum dot}\label{s_imp}
\begin{figure}\begin{centering}
\includegraphics[width=8cm]{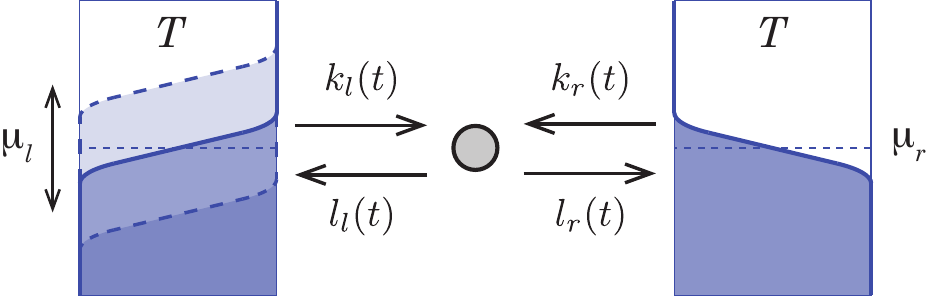}
\caption{Sketch of the system. A single level quantum dot is connected by two reservoirs, allowing the exchange of particles. Indicated in each reservoir is the Fermi distribution and the chemical potentials. Changing the chemical potential in the left reservoir corresponds to a raising/lowering of the distribution compared to the distribution in the right reservoir. The temperatures in both reservoirs are equal.\label{fig2}}
\end{centering}\end{figure}

We now apply the concept of impedance to stochastic transport. We do this by considering one of the simplest stochastic systems that allows a flow of particles: a single quantum dot in simultaneous contact with two electron reservoirs, see Fig.~\ref{fig2}. A difference in chemical potential $\mu_l-\mu_r=qV$ leads to a flow of electrons via the intermediate quantum dot. We restrict ourselves to a quantum dot with only one active level at energy $E$ which is occupied by at most one electron. Hence the quantum dot is either empty (unoccupied), with probability $p_0(t)$, or occupied with probability $p_1(t)$. Changes in the state are due to the exchange of electrons with the two reservoirs, and are described by the following master equation
\begin{equation}
    \dot{\mathbf{p}}(t)=\left(\mathbf{W}^{(l)}(t)+\mathbf{W}^{(r)}(t)\right)\mathbf{p}(t),\label{meq}
\end{equation}
with $\mathbf{p}(t)=(p_0(t),p_1(t))$, and $\mathbf{W}^{(l/r)}(t)$ is the time dependent transition matrix associated with the left/right reservoir
\begin{equation}
    \mathbf{W}^{(l)}(t)=\left(\begin{array}{cc}
       -k_l(t)  &l_l(t)   \\
        k_l(t)  & -l_l(t)
    \end{array}\right),
\end{equation}
and a similar expression for $\mathbf{W}^{(r)}(t)$.
Periodic driving is achieved by time dependent chemical potentials: $\mu_r(t)$ for the right reservoir and $\mu_l(t)$ for the left reservoir. The temperature is constant and for notational simplicity, we set $k_BT=1$. The transition rates satisfy the local detailed balance condition
\begin{equation}
    k_l(t)=e^{-(E-\mu_l(t))}l_l(t)\qquad; \qquad k_r(t)=e^{-(E-\mu_r(t))}l_r(t).\label{detbal}
\end{equation}
This ensures that when the quantum dot is connected to only one reservoir at a constant chemical potential, the time evolution eventually leads to thermal equilibrium.

Unlike the electrical setup, in which there is a single unique electrical current, there are now two possible flows: $j_l$ describing the flow between the left reservoir and the quantum dot, and $j_r$ describing the flow between the quantum dot and the right reservoir:
\begin{equation}
    j_l=k_l(t)p_0(t)-l_l(t)p_1(t)\qquad; \qquad  j_r=l_r(t)p_1(t)-k_r(t)p_0(t)
\end{equation}
In a stationary non equilibrium situation, with constant but different chemical potentials, these flows are identical, but in case of time dependent chemical potentials, this is no longer the case.

\begin{figure}\begin{centering}
\includegraphics[width=12cm]{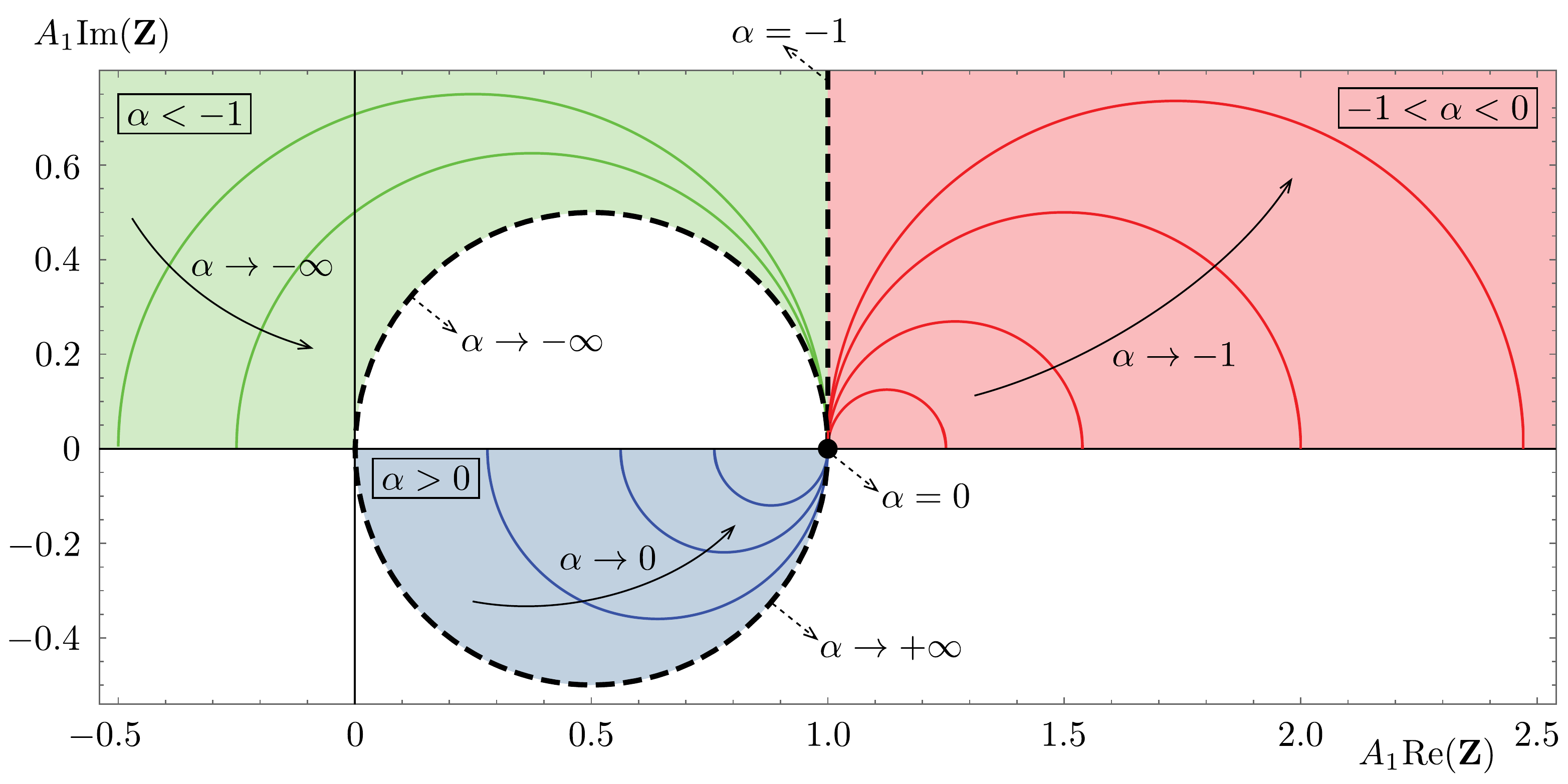}
\caption{Plot of the imaginary and real part of the impedance $\z{Z}$ given by Eq.~(\ref{ZA}) as a function of the ratio $\alpha$. The three regions (see text) are indicated by the boxed interval of $\alpha$ values and by the different colours. Solid arrows show the direction of variation in $\alpha$, dashed arrows indicate specific values of $\alpha$.\label{fig3}}
\end{centering}\end{figure}

In the context of impedance, we consider a time periodic driving. In that situation, eventually the flows and probabilities will become time periodic. Eq.~(\ref{meq}) can be solved exactly, and gives (after short-time initial-state corrections have vanished) \cite{risken1996fokker}:
\begin{equation}
    \mathbf{p}(t)=\mathbf{p}_{\textrm{\tiny{ad}}}(t)-\int^{\infty}_0d\tau e^{-\int^{\tau}_0ds\left[k(t-s)+l(t-s)\right]}\dot{\mathbf{p}}_{\textrm{\tiny{ad}}}(t-\tau),
\end{equation}
with 
\begin{equation}
    \mathbf{p}_{\textrm{\tiny{ad}}}(t)=\frac{1}{k(t)+l(t)}\left(\begin{array}{c}
         l(t)  \\
         k(t)
    \end{array}\right),
\end{equation}
the steady-state distribution if the driving was fixed at time $t$, and
\begin{equation}
k(t)=k_l(t)+k_r(t) \;\;\;\mbox{and}\;\;\;l(t)=l_l(t)+l_r(t).
\end{equation}
The flow from the left reservoir into the quantum dot then reads:
\begin{eqnarray}
    j_l&=&\frac{k_l(t)l_r(t)-l_l(t)k_r(t)}{k(t)+l(t)}\nonumber\\&&+\int^{\infty}_0d\tau e^{-\int^{\tau}_0ds\left(k(t-s)+l(t-s)\right)}\big(k_l(t)+l_l(t)\big)\nonumber\\
    &&\hspace{2cm}\times \left(\frac{\dot{k}(t-\tau)l(t-\tau)-k(t-\tau)\dot{l}(t-\tau)}{(k(t-\tau)+l(t-\tau))^2}\right),\label{jdef}
\end{eqnarray}
and a similar expression for $j_r$. As these expressions depend nonlinearly on the chemical potentials, we first need to linearise them. Without loss of generality, we set the chemical potential of the right reservoir constant, $\mu_r(t)=\mu_0$, and the driving $\Delta \mu(t)$ of the left reservoir small enough $\mu_l(t)=\mu_0+\Delta \mu(t)$. Expanding $k_l(t)$ and $l_l(t)$ to first order in $\Delta \mu(t)$
\begin{equation}
l_l(t)\approx l_{l;0}+\Delta \mu(t) l_{l;1}\quad\quad ; \quad\quad k_l(t)\approx k_{l;0}+\Delta \mu(t) k_{l;1},
\end{equation}
while the rates associated with right reservoir are constant, $k_r(t)=k_{r;0}$ and $l_r(t)=l_{r;0}$.
The local detailed balance condition, Eq.~(\ref{detbal}), leads to the following relation between the coefficients
\begin{eqnarray}
    \frac{k_{l;0}}{l_{l;0}}= \frac{k_{r;0}}{l_{r;0}}\quad\quad ;\quad\quad k_{l;1}=k_{l;0}+\frac{k_{l;0} l_{l;1}}{l_{l;0}}.
\end{eqnarray}
Linearising Eq.~(\ref{jdef}) in terms of the driving $\Delta \mu(t)$ gives,
\begin{equation}\label{jl_lin}
    j_l=\frac{\Delta \mu(t) k_{l;0}l_{r;0}}{k_0+l_0}+\frac{k_{l;0}l_0(k_{l;0}+l_{l;0})}{(k_{0}+l_{0})^2}\int^{\infty}_0d\tau e^{-(k_0+l_0)\tau}\Delta\dot{\mu}(t-\tau),
\end{equation}
with $k_0=k_{l;0}+k_{r;0}$, $l_0=l_{l;0}+l_{r;0}$. The corresponding result for $j_r$ reads:
\begin{equation}\label{jr_lin}
       j_r=\frac{\Delta \mu(t) k_{l;0}l_{r;0}}{k_0+l_0}-\frac{k_{l;0}l_0(k_{r;0}+l_{r;0})}{(k_{0}+l_{0})^2}\int^{\infty}_0d\tau e^{-(k_0+l_0)\tau}\Delta\dot{\mu}(t-\tau).
\end{equation}
Similar expressions for the flows can be obtained when considering other configurations, for example $\Delta \mu_l(t)=-\Delta \mu_r(t)=\Delta \mu/2$, which allows us to postulate the following generic expression:
\begin{equation}
j=A_1\Delta \mu(t)+A_2\int_0^{\infty}d\tau\,e^{-(k_0+l_0)\tau}\Delta \dot{\mu}(t-\tau).
\end{equation}
This flow is decomposed in two parts. The first term represents an 'adiabatic' contribution, which is the steady-state flux associated with the gradient $\Delta \mu(t)$ fixed at time $t$. The flow associated with this term is directly proportional to the gradient, and can be considered as an Ohmic contribution. The second term is due to the finite speed of the  driving. This term will lead to a phase difference between the current $j$ and gradient $\Delta \mu$. Introducing $\Delta \mu(t)=\mu e^{i\omega t}$ allows to calculate the impedance $\z{Z}=\Delta \mu(t)/j$ \footnote{In a thermodynamic setting the driving force for a particle current is the difference in chemical potential divided by $kT$.}:
\begin{equation}
\z{Z}=\left(A_1+\frac{i\omega A_2}{k_0+l_0+i\omega}\right)^{-1}\label{ZC}.
\end{equation}
Since the sign of $A_1$ can always be made positive by an appropriate choice for the direction of the current, it is clear from this expression that the qualitative dependence of $\z{Z}$ on the parameters is fully determined by the ratio $\alpha \equiv A_2/A_1$. Introducing $\omega'=\omega/(k_0+l_0)$ leads to
\begin{equation}
    A_1\z{Z}=\frac{1+\omega'^2(1+\alpha)-i\omega'  \alpha}{1+\omega'^2(1+\alpha)^2}.\label{ZA}
\end{equation}
The following result is immediate:
\begin{equation}
\left(\textrm{Re}(A_1\z{Z})-\frac{2+\alpha}{2+2\alpha}\right)^2+\textrm{Im}(A_1\z{Z})^2=\left(\frac{\alpha}{2+2\alpha}\right)^2.
\end{equation}
That is, the real and imaginary part are always located on a circle with the centre on the real axis. In fact, when the frequency is varied from 0 to $\infty$,  $\textrm{Re}(\z{Z})$ and $\textrm{Im}(\z{Z})$ always trace out a semicircle, since $\textrm{Im}(\z{Z})$ does not changes sign, and starts/ends in 0. The analogy with the electrical circuit shown in Fig.~\ref{fig1} with $\z{Z}_D\equiv R_D$ follows immediate. A direct comparison with Eq.~(\ref{zRD}) gives the following identification
\begin{equation}
R=\frac{1}{A_1(1+\alpha)}\quad ; \quad R_D=\frac{\alpha}{A_1(1+\alpha)}\quad ; \quad C=\frac{A_1(1+\alpha)^2}{\alpha(k_0+l_0)}.
\end{equation}
While the analogy is there, the interpretation of such an identification is not straightforward. In fact, depending on the value of $\alpha$, the signs of $R$, $R_D$ and $C$ can change. These changes in sign can be used to identify three different regions in the $\left(\textrm{Re}(\z{Z}),\textrm{Im}(\z{Z})\right)$-plane by varying $\alpha$ from $-\infty$ to $+\infty$. A graphical representation of Eq.~(\ref{ZA}) is given in Fig.~\ref{fig3}. The first region, for positive $\alpha$ and shown as the blue region in Fig.~\ref{fig3}, corresponds to positive values for $R$, $R_D$ and $C$. When $\alpha=0$ (or $A_2=0$) the current only contains the adiabatic contribution. Hence $\textrm{Im}(\z{Z})=0$ and the radius of the semi-circle reduces to zero. This result is equivalent with an electrical circuit containing a single resistor $R$. As $\alpha$ decreases further, the values for $R$, $R_D$ and $C$ can become negative. For $-1 < \alpha < 0$ (the red region in Fig.~\ref{fig3}) we find $R>0$ and both $R_D$ and $C$ negative (see for example \cite{jonscher1986,datta2008}). This region ends at $\alpha=-1$. For that specific value the radius of the semi-circle diverges and becomes a straight vertical line. This impedance is equivalent to that of a series combination of a resistor and inductor, as Eq.~(\ref{ZC}) reduces to
\begin{equation}
\z{Z}=\frac{1}{A_1}+\frac{i\omega}{A_1(k_0+l_0)}\label{ZL}.
\end{equation}
Finally, the third region, indicated by the green color in Fig.~\ref{fig3}, corresponds to $\alpha < -1$. In this case $R<0$, $R_D>0$ and $C<0$ (the green region in Fig.~\ref{fig3}).

These results show that the impedance in a stochastic setup, even for a simple system as considered here, can be quite diverse. The qualitative behaviour strongly depends on the values of the various parameters. Unlike the electrical setup, the characteristics of the impedance are not due to the presence of different components. In contrast, here they have a dynamic origin and are due to the difference in time scales of the driving frequency $\omega$ and the stochastic events as determined by the transition rates. 

So far the calculations were done for general transition rates satisfying the detailed balance condition, without further assumptions. A specific choice is made by considering the connected thermal reservoirs as metallic leads described by a Fermi-Dirac distribution (see for example \cite{rutten2009}), leading to:
\begin{eqnarray}
k_l(t)&=\Gamma_l f(x_l) \;\;\;;\;\;\; l_l(t)=\Gamma_l \left(1-f(x_l)\right)\\
k_r(t)&=\Gamma_r f(x_r) \;\;\;;\;\;\; l_r(t)=\Gamma_r \left(1-f(x_r)\right)
\end{eqnarray}
with
\begin{equation}
f(x)=\frac{1}{e^{x}+1} \;\;\;;\;\;\; x_l=\frac{E-\mu_l(t)}{kT}\;\;\;;\;\;\; x_r=\frac{E-\mu_r(t)}{kT}.
\end{equation}
The presence of $E$ is only visible in the end results via a prefactor in the currents, hence without loss of generality, we can set $E=0$ and (as before) $kT=1$. Further setting $\mu_r(t)=\mu_0=0$ and $\mu_l(t)=\mu e^{i\omega t}$ we end up with the following (linearised) currents:
\begin{equation}
j_l=\frac{\mu e^{i\omega t}\Gamma_l(\Gamma_r + i\omega)}{4\left(\Gamma_l+\Gamma_r + i\omega\right)}
\end{equation}
and
\begin{equation}
j_r=\frac{\mu e^{i\omega t}\Gamma_l\Gamma_r}{4\left(\Gamma_l+\Gamma_r + i\omega\right)}.
\end{equation}
The corresponding impedances are
\begin{equation}
\z{Z}_l=\frac{4}{\Gamma_l}+\frac{4\Gamma_r}{\Gamma_r^2 + \omega^2}-i\frac{4\omega}{\Gamma_r^2 + \omega^2}
\end{equation}
and
\begin{equation}
\z{Z}_r=\frac{4}{\Gamma_l}+\frac{4}{\Gamma_r}+i\frac{4\omega}{\Gamma_l\Gamma_r}.
\end{equation}
These results show that it is not possible to assign a unique impedance to a stochastic system. Unlike the electrical counterpart, the current here depends on the location at which it is measured. A calculation of the impedance based upon either $j_l$ or $j_r$ yields quite a different result. Comparing $\z{Z}_l$ with Eq.~(\ref{zRD}) shows that this impedance is located in the blue region ($\alpha >0$) and hence is equivalent to the circuit to the electrical circuit shown in Fig.~\ref{fig1} with $\z{Z}_D\equiv R_D$. $\z{Z}_r$ on the other hand corresponds to the vertical dashed line in Fig.~\ref{fig3} with $\alpha=-1$.

%%%%%%%%%%%%%%%%%%%%%%%%%%%%%%%%%%%%%%%%%%% 
\section{Conclusions}\label{outlook}
In conclusion, we have introduced and applied the concept of impedance in a stochastic setting, namely the stochastic flow of particles. For a single-level quantum dot connecting two electron reservoirs, the impedance can be calculated exactly. Even this seemingly simple setup displays a diverse response to an alternating driving. For certain parameter values, the impedance is equivalent to that of an actual electrical circuit. An interesting question for further research would be how general this result is, i.e., to what extend is it possible to map a stochastic system on an equivalent electrical circuit and vice versa and whether a general procedure exists to do this mapping. This might be done using general approaches such as macroscopic fluctuation theory \cite{derrida2007non,bertini2015macroscopic}.

Another open question, is whether an actual impedance measurement on an experimental system (such as nanowires or the cable bacteria mentioned in the introduction) shows any signs of the underlying transport mechanism. This is clearly of interest, as it leads to new insights concerning the internal structures of the experimental system at hand. Results obtained so far for cable bacteria only show a purely resistive behaviour \cite{cbRobin} up to frequencies of 1MHz. Comparing this frequency with the typical hopping rates applicable in bacterial nanowires, which are of the order of $10^{13}$ $s^{-1}$ (see for example \cite{C1FD00098E}), shows that these driving frequencies are rather low. And as a result, the imaginary part of the impedance, responsible for the capacitive signature, becomes insignificant.

Apart from the biological relevance, stochastic impedance also has a fundamental appeal. The field of thermodynamics has made tremendous progress in the last decades with the development of stochastic thermodynamics \cite{seifert2012stochastic,van2015ensemble}. In particular, the analysis done in this work bears similarities with the framework for linear stochastic thermodynamics for periodically driven systems \cite{brandner2015thermodynamics,proesmans2015onsager,proesmansJSTAT2016,brandner2016periodic}. In these works, the thermodynamic quantities such as for example heat, (chemical) work and entropy have been defined as time-averages over one period of the driving signal. The use of impedance allows to describe these quantities in a fully time dependent setting, and might reveal new characteristics concerning the thermodynamics of small-scaled systems.

%%%%%%%%%%%%%%%%%%%%%%%%%%%%%%%%%%%%%%%%%%% 
\section*{Acknowledgements}
K. P. is a postdoctoral fellow of the Research Foundation-Flanders (FWO) under Grant n$^{\circ}$ 12J2819N. Valuable discussions and feedback from X-LAB, Ralf Eichhorn and Carlos E. Fiore are gratefully acknowledged.
%%%%%%%%%%%%%%%%%%%%%%%%%%%%%%%%%%%%%%%%%%%% 

\end{document}